# Fast and Chaotic Fiber-Based Nonlinear Polarization Scrambler

M. Guasoni, P-Y. Bony, M. Gilles, A. Picozzi, and J. Fatome

*Abstract*— We report a simple and efficient all-optical polarization scrambler based on the nonlinear interaction in an optical fiber between a signal beam and its backward replica which is generated and amplified by a reflective loop. When the amplification factor exceeds a certain threshold, the system exhibits a chaotic regime in which the evolution of the output polarization state of the signal becomes temporally chaotic and scrambled all over the surface of the Poincaré sphere. We derive some analytical estimations for the scrambling performances of our device which are well confirmed by the experimental results. The polarization scrambler has been successfully tested on a single channel 10-Gbit/s On/Off Keying Telecom signal, reaching scrambling speeds up to 250-krad/s, as well as in a wavelength division multiplexing configuration. A different configuration based on a sequent cascade of polarization scramblers is also discussed numerically, which leads to an increase of the scrambling performances.

## I. Introduction

The ability to randomly scramble the state-of-polarization (SOP) of a light beam is an important issue that encounters numerous applications in photonics. Polarization scrambling is indeed mainly implemented to ensure polarization diversity in optical telecommunication systems so as to combat deleterious polarization effects and provide mitigation of polarization mode dispersion (PMD) and polarization dependent loss or gain [1]. For instance, polarization scrambling has been exploited to avoid polarization hole burning in Erbium doped fiber amplifiers (EDFA) [2], and has allowed washing out PMD-induced error bursts within forward error correction frames [3]. Furthermore, polarization scrambling is a mandatory procedure when testing the performances of polarization-sensitive fiber systems or optical components. For that purpose, the SOP changing rate (i.e. the scrambling speed) induced by the scrambler device should be as high as some hundreds of Krad/s in order to match the scale of fast polarization changes encountered in high-speed fiber optic systems [4].

Traditionally, polarization scrambler technology is based on the cascade of fiber resonant coils, of rotating half and quarter alternated wave-plates, or of fiber squeezers as well as opto-electronic elements [5-11]. In most of these devices, an external voltage is applied: it drives the rotation of the wave-plates, the squeezing of the fiber as well as the expansion of the piezo-electric coils, so that the scrambling performances are directly controlled by means of this driving voltage. Thanks to such opto-electronic technologies, records of scrambling speeds have been reported, reaching several of Mrad/s [10, 11]. Nevertheless, one can remark that these current commercially available solutions exhibit the common property of being essentially deterministic methods. Indeed, these devices impose to an incident light beam repetitive trajectories on the surface of the Poincaré sphere, which could be seen as a limitation when one would mimic true in-field optical links which are known to exhibit a stochastic dynamics.

The aim of this work is to report a theoretical and experimental description of an all-optical, fully chaotic polarization scrambler, which is shown to exhibit a genuine chaotic dynamics. Furthermore, the present device could be qualified as "home-made" since it is essentially based on standard components usually available in any Labs working in the field of nonlinear optics and optical communications.

The basic principle of this device was first established in ref. [12] in order to demonstrate a transparent method of temporal spying and concealing process for optical communications. Indeed, it consists in an additional operating mode, namely the chaotic mode, of the device called Omnipolarizer, originally conceived to operate as an all-optical polarization attractor and beam splitter [13, 14].

In this paper we gain a deeper insight into the physics of this all-optical polarization scrambler. First in section II, we introduce the principle of operation of our polarization scrambler. Then in section III we describe the experimental setup. In section IV we develop our theoretical modeling and derive an analytical estimation for two important threshold parameters, which allow us to carefully discuss the transition of the system from a polarization attraction regime to the chaotic scrambling regime. In section V we report our experimental results in the CW regime and for a 10-Gbit/s On/Off keying (OOK) Telecom signal. Then, in section VI, we find an estimation of the scrambling performances as a function of the system parameters, and in section VII we discuss the efficiency improvement provided by a cascade of scramblers. In section VIII we provide experimental evidences of the compatibility of our all-optical scrambler for WDM applications and in the last section we trace out the conclusions.

## II. PRINCIPLE OF OPERATION

The principle of the proposed all-optical scrambler is schematically displayed in Fig. 1. It basically consists in a nonlinear Kerr medium, here an optical fiber, in which an initial forward signal **S** with a fixed polarization-state nonlinearly interacts through a cross-polarization interaction with its own backward replica **J** generated and amplified at the fiber end by means of a reflective loop. A strong power imbalance between the two beams is applied, which can actually lead to a chaotic polarization dynamics of both the forward and backward output fields [15, 16].

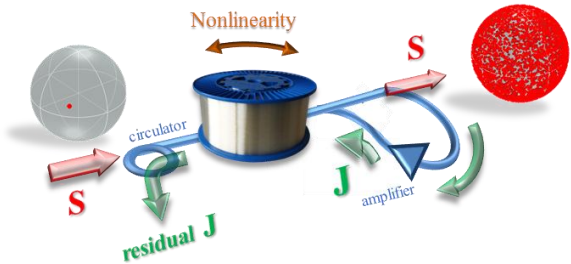

Fig. 1. Principle of operation.

In some of our previous works [13, 14, 17, 18] we have already identified some particular regimes associated to this kind of counter-propagative cross-polarization interactions. For example, we have put in evidence that typically for nearly equal forward and backward beam powers, the stable stationary singular states of the system play the role of natural polarization attractors for the output signals. Generally these stationary states **S**$_{stat}$ and **J**$_{stat}$ can be computed as a function of the system parameters, such as the forward and backward powers and the fiber length $L$.

In order to illustrate that point, panels (a-c) of Fig. 2 display the attraction process undergone by an input signal towards a stable stationary state for nearly equal counter-propagative beam powers. For clearness, only a single Stokes component of **S** and **J** is represented, let us say $S_1$ and $J_1$ (solid line), as well as for **S**$_{stat}$ and **J**$_{stat}$, let us say $S_{1,stat}$ and $J_{1,stat}$ (in circles). We can then observe the spatial evolutions of $S_1$ and $J_1$ along the fiber length for 3 consecutive times $t_A<t_B<t_C$. In the time slot $0<t<L/c$, where $c$ is the speed of light in the fiber, the signal **S** propagates unaffected by nonlinear effects, since **J** has not been yet generated at fiber end. For illustration purpose, panel (a) shows the corresponding spatial profile of $S_1$ at instant $t_A$ slightly larger than $L/c$: the backward replica $J_1$ has just been reflected and begins to counter-propagate. Afterward, (panel (b)), the cross-polarization interaction between **S** and **J** makes them to gradually converge towards the stable stationary states of the system. Finally, at the instant $t_C$ (panel (c)) the spatial profiles of **S** and **J** almost perfectly match the stationary solutions and do not evolve substantially in the subsequent instants. Therefore, in this instance the stable stationary states act as asymptotic attractors [14].

On the contrary, for large power unbalances between counter-propagating fields, the stationary states become unstable. As depicted by panels (d-f) the forward and backward beams are then no longer attracted towards a stationary state in the time solution: both beams oscillate in time without reaching a fixed state. We will see in the following that the forward polarization at the exit of the fiber varies endlessly in time and becomes temporally scrambled as a result of a chaotic dynamics all over the surface of the Poincaré sphere. This constitutes the basic principle underlying the operation of our all-optical polarization scrambler.

## III. EXPERIMENTAL IMPLEMENTATION

The experimental implementation of the proposed scrambler is schematically displayed in Fig. 3.

For fundamental studies, the initial signal consists in a fully polarized 100-GHz-bandwidth partially incoherent wave, centered at 1550-nm. This incident signal is generated from an Erbium-based amplified spontaneous noise source (ASE) sliced in the spectrum domain by means of a wavelength-demultiplexer followed by an inline polarizer. This large bandwidth input signal is used to avoid any impairment due to the stimulated Brillouin backscattering in the fiber under-test.

In a second step, in order to evaluate the performance of this all-optical scrambler for Telecom applications, the incoherent wave is replaced by a 10-Gbit/s OOK signal at 1550-nm. This return-to-zero (RZ) optical signal is generated from a 10-GHz mode-locked fiber laser delivering 2.5-ps pulses at 1550-nm. The spectrum of this initial pulse train is sliced thanks to a liquid-crystal based optical filter to broad the pulses to 20-ps. The resulting 10-GHz pulse train is intensity modulated thanks to a LiNbO$_3$-based Mach-Zehnder modulator driven by a high-speed RF pattern generator. The input signal is then amplified by means of an Erbium doped fiber amplifier (EDFA-1) before injection into the fiber under-test thanks to an optical circulator. Note that this optical circulator is mainly used so as to evacuate the residual counter-propagating signal replica.

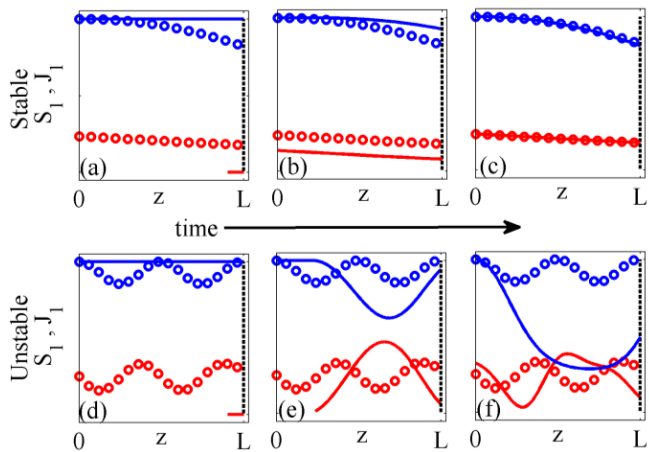

Fig. 2. Spatial evolution along the fiber length of the Stokes components $S_1$ (blue solid lines) and $J_1$ (red solid lines) at 3 consecutive instants $t_A$ (a-d), $t_B$ (b-e) and $t_C$ (c-f). Corresponding stationary solutions $S_{1,stat}$ and $J_{1,stat}$ are represented in circles. In the case of panels (a-c) the counter-propagating waves have almost the same power so that the stationary solutions are stable. Consequently $S_1$ and $J_1$ gradually converge in time towards $S_{1,stat}$ and $J_{1,stat}$ respectively. Conversely, in the case depicted by panels (d-f) the stationary solutions are unstable and therefore no attraction process occurs.

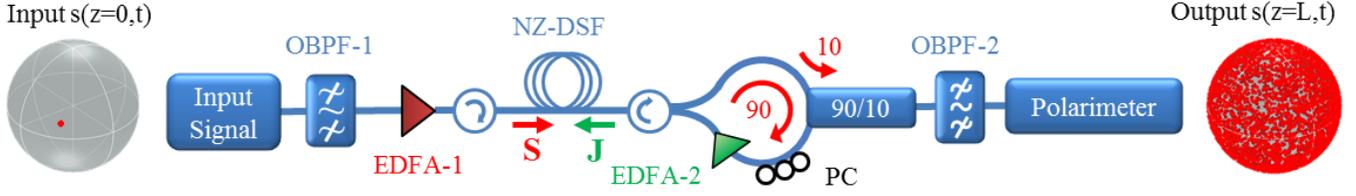

Fig. 3 Experimental setup of the chaotic polarization scrambler under study. PC: Polarization Controller, OBPF: Optical bandpass filter. The two Poincaré spheres illustrate the distribution of the Stokes vector of the forward beam at the input (fully polarized) and at the output (scrambled over the sphere), respectively.

In order to characterize the dynamics of the polarization scrambling regime, two fibers with different lengths were tested. Fibre-1 corresponds to a L=5.3 km non-zero dispersion shifted fiber (NZ-DSF) characterized by a chromatic dispersion D= -1 ps/nm/km at 1550 nm, linear losses $\alpha$=0.24 dB/km and a nonlinear coefficient $\gamma$=1.7 $W^{-1}km^{-1}$. Fibre-2 is a highly normal dispersive fiber (OFS-HD) characterized by L=10 km, D=-14.9 ps/nm/km, $\gamma$=2.3 $W^{-1}km^{-1}$ and losses $\alpha$=0.2 dB/km.

The reflective-loop generating the backward replica at the fiber end consists in an optical circulator followed by a polarization controller (PC) and Erbium amplifier (EDFA-2). The EDFA-2 provides the amplification of the backward beam, which allows controlling the power imbalance with respect to the forward signal. A 90/10 coupler is also inserted into the loop to drive the resulting scrambled signal for analysis. The output forward signal is then optically filtered at $\lambda_{OBPF-2}$=1550 nm (OBPF-2, bandwidth 100-GHz) to suppress the excess of amplified spontaneous noise emission outside the signal bandwidth. The state-of-polarization of the output signal is finally characterized by means of a commercial polarimeter unit.

## IV. THEORETICAL MODELING

In the following, we indicate with $S=[S_1,S_2,S_3]$ and $J=[J_1,J_2,J_3]$ the Stokes vectors for the forward and backward beams, respectively. Consequently, the normalized unitary vectors $s=S/|S|$ and $j=J/|J|$ indicate the corresponding SOP.

The dynamics of the system is mainly driven by the amplification factor $g$ of the loop defined by the power ratio between the backward and forward signals at the fiber exit: $g=|J(z=L,t)|/|S(z=L,t)|$, where $z$ indicates the propagation length along the fiber. In practice, the coefficient $g$ is directly controlled by means of the EDFA-2. Furthermore, in ref. [12], we have pointed out the existence of two threshold values for $g$, defined as $g_A$ and $g_C$, that allow us to distinguish 3 different kinds of regimes of operation, namely the *attraction* regime ($g<g_A$), the *transient* regime ($g_A<g<g_C$) and the *chaotic* regime ($g>g_C$).

The evolutions of $S$ and $J$ in the fiber are governed by the following coupled equations [19]:

$$c^{-1}\partial_t S + \partial_z S = S \times DJ - \alpha S \\ c^{-1}\partial_t J - \partial_z J = J \times DS - \alpha J \quad (1)$$

where D=$\gamma$diag(-8/9,8/9,-8/9) is a diagonal matrix, $\gamma$ and $\alpha$ are the nonlinear Kerr coefficient and the propagation losses of the fiber, respectively, and $c$ is the speed of light in the fiber. According to Eqs.(1) the average powers $P_S(z)=\langle|S(z,t)|\rangle$ and $P_J(z)=\langle|J(z,t)|\rangle$ (the brackets $\langle\rangle$ denote a temporal averaging) are individually conserved except for the propagation losses, indeed $P_S(z)=P_S(0)\exp(-\alpha z)$ and $P_J(z)=P_J(L)\exp(\alpha(z-L))$.

In our numerical simulations we solve Eqs.(1) subject to the boundary condition $J(z=L,t)=gRS(z=L,t)$, in which $R$ is a 3x3 matrix modeling the polarization rotation in the reflective-loop, which is imposed by the circulator and adjusted by means of the polarization controller (PC). The matrix $R$ is defined by $R=R_x(\theta)R_y(\beta)R_z(\chi)$, being $R_{x,y,z}$ three standard rotation matrices and $\theta, \beta, \chi$ the corresponding rotation angles around the $x$, $y$ and $z$ axes of the Poincaré sphere, respectively.

In the configuration under-study, the dynamics of $S$ and $J$ are related to the stability of the stationary states of the system, which are the solutions of Eqs.(1) in the CW limit, i.e. when dropping the time derivatives.

In the limit where losses are neglected the stationary states of Eqs.(1) read as [20]:

$$S(z) = [\,S(0) - \Omega \bullet S(0)\,\Omega/|\Omega|^2\,]\cos(|\Omega|z) \\ + \Omega \bullet S(0)\,\Omega/|\Omega|^2 + [\,\Omega \times S(0)/|\Omega|\,]\sin(|\Omega|z) \quad (2)$$

where $\bullet$ indicates the scalar product and $\Omega=S-DJ$ is an invariant throughout the fiber. In [20] the sequent relation is reported which ties the input polarization alignment $\mu=K^{-1}[-DJ(L)]\bullet S(0)$ and the output polarization alignment $\eta=K^{-1}[-DJ(L)]\bullet S(L)$, being $K=|DJ||S|$ a system invariant:

$$\mu = (|DJ|\eta+|S|)(|S|\eta+|DJ|)(1-\cos(|\Omega|L)) \\ /(|DJ|^2+|S|^2+2K\eta)+\eta\cos(|\Omega|L) \quad (3)$$

Here we underline that, since $|DJ| \equiv \gamma(8/9)|J| \equiv \gamma(8/9)|gRS| \equiv \gamma(8/9)g|S|$, then in the limit $g\gg1$, Eqs.(3) gives $\mu\simeq\eta$, which gives:

$$s(L) \bullet DR\,s(L) = s(0) \bullet DR\,s(L) \quad (4)$$

The ensemble of the output stationary SOPs **s**(*L*) solving Eq.(4) describes a closed line over the Poincaré sphere, which we call here Line of Stationary Output SOPs (LSOS) and whose shape depends both on the input **s**(0) and on the rotation matrix *R*.

In ref. [17] a general rule is reported concerning the stability of these stationary solutions, stating that a stationary solution is stable if it exhibits a non-oscillatory evolution along the whole fiber length. We point out that the vector **S**(*z*) given by Eq.(2) is formed by the three orthogonal components **Ω**·**S**(0) **Ω** / |**Ω**|$^2$ , [ **S**(0) − **Ω**·**S**(0) **Ω** / |**Ω**|$^2$ ] cos(|**Ω**|*z* ) and [**Ω** ×**S**(0)/ |**Ω**|] sin(|**Ω**|*z*) that are all monotonic in *z* if |**Ω**|*L*<π/2. Considering that if *g* >>1 then |**Ω**| ≡ |**S** − **DJ**| ≃ |**DJ**| ≡ *γ*(8/9)*g*|**S**|, we obtain that a stationary state is stable only if the condition *γg*|**S**|*L*<9π/16 is satisfied. We remind that this condition holds in the limit where losses can be neglected. Note however that our numerical simulations confirm that this conditions still holds in presence of reasonable small losses (typically 0.2 dB/km) after substitution of |**S**| with |**S**(*L*)|≡ *P*$_S$(*L*). Actually if *γgP*$_S$(*L*)*L*<9π/16, or equivalently *g*<9π/(16·*L*·*Ps*(*L*)·*γ*), then the stationary states belonging to the LSOS are stable and represent an attraction point over the Poincaré sphere. We thus confirm the existence of an attraction regime that, as already observed in ref. [12], is characterized by an upper threshold *g*$_A$ here estimated by the relation *g*$_A$=9π / (16·*L*·*Ps*(*L*)·*γ*). In this regime if a CW SOP **s**(0,*t*)=**s**(0) is injected in the fiber then the corresponding output SOP **s**(*L*,*t*) always converges in time towards a fixed point belonging to the LSOS, which is analogous to the attraction process experienced by $S_1$(L,t) in Fig.2(a-c).

The position of the point over the LSOS depends on |**Ω**|*L*, and thus on the product *γgP*$_S$(*L*). This means that, by varying the value of the amplification factor *g*, different points over the LSOS can be reached.

On the other hand, when *g*>*g*$_A$ the system no longer operates in the attraction regime. More precisely, a threshold *g*$_C$ is found such that, if *g*$_A$<*g*<*g*$_C$ the system operates in a transition regime where the output SOP could reach a constant-in-time value, as well as a periodic, or even a chaotic temporal trajectory. The type of dynamics of the system is shown to depend on the particular input SOP and on the particular rotation matrix R. Finally, when *g*>*g*$_C$, a chaotic regime all over the surface of the Poincaré sphere is reached, irrespective of the input SOP and the rotation matrix R. It is the operation regime which underlies the basic principle of the proposed polarization scrambling device. Note that, as will be discussed later, this chaotic dynamics is characterized by the presence of a positive Lyapunov coefficient [12].

Our numerical simulations show that the threshold gain *g*$_C$ beyond which a chaotic regime occurs is typically in the range of [5-10] *g*$_A$. Both *g*$_A$ and *g*$_C$ are thus ∝ (*L*·*Ps*(*L*))$^{-1}$≡(*L*·*Ps*(0)·10$^{-\alpha L/10}$)$^{-1}$, therefore for typical propagation losses of about 0.2 dB/km and a fiber length *L*<20 km these thresholds can be reduced by increasing the fiber length.

Let us now illustrate the general phenomenology of the dynamics of the system by considering numerical simulations with the experimental parameters of Fiber-1. The results are reported in Figs. 4 when a P$_S$(0)=15 dBm CW forward signal is injected into the system.

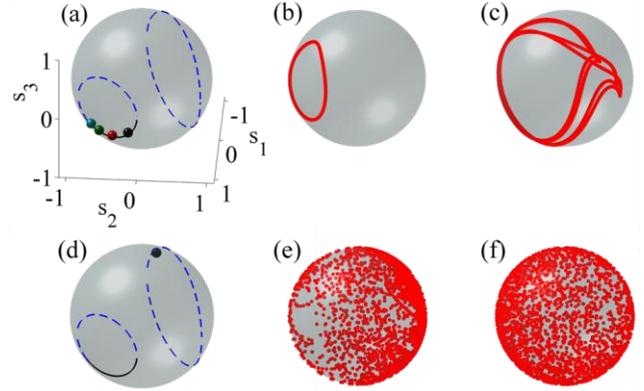

Fig.4 Output distribution of the SOP **s**(*L*,*t*) over the Poincaré sphere for increasing values of the reflective coefficient *g* (see text for details about the system parameters). Panel (a): fixed stable points reached by **s**(*L*,*t*) when *g*=2 (black dot), *g*=4 (red dot), *g*=6 (green dot) and *g*=8 (cyan dot). Panels (b) and (c): periodic trajectories corresponding to *g*=12 and g=16, respectively. Panel (d): fixed unstable point reached by **s**(*L*,*t*) when *g*=25. Panel (e): semi-chaotic trajectory corresponding to *g*=28. Panel (f): chaotic trajectory corresponding to *g*=50. The closed curves in panels (a, d) form the Line of Stationary Output SOPs (LSOS), which is defined by the equations {2·$s_1$·$s_2$ + $s_3^2$ = $s_2$; $s_1^2$+ $s_2^2$+ $s_3^2$=1}. The black solid line corresponds to the stable part of the LSOS, composed by the stable stationary states; the blue dotted line corresponds to the unstable part, composed by the unstable stationary states.

For this configuration we have P$_S$(L)=13.8 dBm, and we can thus estimate the threshold gain *g*$_A$ = 9π/(16·L·Ps(L)·γ) ≃ 8 as well as the threshold *g*$_C$ ≃ 5*g*$_A$ ≡ 40. For this series of simulations, the SOP of the input signal is aligned with the x-axis of the Poincaré sphere, that is **s**(0,*t*) = **s**(0) = (1,0,0), and the rotation angles are *θ*=0, *β*=0 and *χ*=π/2, that is R=[(0,1,0); (-1,0,0); (0,0,1)]. In this case, Eq.(4) reads as 2·$s_{L1}$·$s_{L2}$ + $s_{L3}^2$ = $s_{L2}$, where $s_{L1,2,3}$ are the components of **s**(*L*) and are subject to the constraint |**s**|$^2$ ≡ $s_{L1}^2$+$s_{L2}^2$+$s_{L3}^2$ = 1. The corresponding LSOS, formed by two closed and distinct curves over the Poincaré sphere, are plotted in Fig. 4a, where both the stable (black solid line) and the unstable part (blue dashed line) of the LSOS are put in evidence.

Fig. 4a illustrates the attraction regime, i.e. when *g*<*g*$_A$. It displays the fixed points that are reached by the output SOP when *g*=2 (black dot), *g*=4 (red dot), *g*=6 (green dot) and *g*=8 (cyan dot), respectively. As predicted theoretically, a unique deterministic point is reached for each value of *g* and more importantly, such point lies over the stable part of the LSOS.

Figures 4(b-e) illustrate the transition regime, that is to say when *g*$_A$<*g*<*g*$_C$. As previously mentioned, more or less complex periodic trajectories can be observed in this regime, for instance in panel (b) for *g*=12 and in panel (c) for *g*=16, as well as fixed unstable points (panel (d), g=25). Indeed, our numerical simulations reveal that if a fixed point is reached in the transition regime, then it always belongs to the unstable part of the LSOS. For this reason even a small perturbation of the system parameters, for example of the coefficient $g$ or of the rotation matrix $R$, leads to a dramatic change of the output dynamics of $s(L,t)$, which can evolve towards a complex periodic or semi-chaotic trajectory. This feature is clearly visualized in panels (d) and (e), corresponding to a variation of the amplification factor from $g=25$ to $g=28$.

The semi-chaotic trajectory in Fig. 4(e) is the signature of the passage from the transition regime to the chaotic regime: the path of $s(L,t)$ over the Poincaré sphere exhibits an apparent random motion, although it only fills a part of the surface of the Poincaré sphere. In the chaotic regime ($g>g_C$, panel (f)) the trajectory is uniformly distributed all over the surface of the Poincaré sphere, so that an efficient and nondeterministic polarization scrambling of the output signal is achieved, in agreement with our theoretical predictions. Obviously, this is the ideal operation regime for the present all-optical polarization scrambler.

## V. EXPERIMENTAL RESULTS

In order to confirm the theoretical and numerical predictions, a series of experiments have been carried out by considering the Fiber-1 configuration. A 100-GHz incoherent signal was injected into the system with a fixed and arbitrary polarization state as well as a constant power $P_S(0)=15$ dBm.

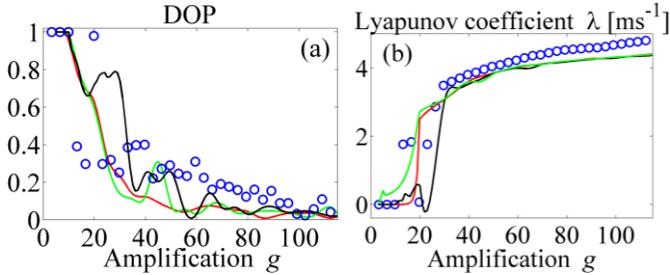

Fig. 5 DOP of the output signal as a function of the reflective coefficient $g$, i.e. the backward power. The blue circles correspond to the experimental measurements. The DOP close to the unity around $g=20$ corresponds to the attraction towards an unstable stationary states in the transient regime. Color solid lines display the numerical simulations with 3 particular different rotation matrices R and 3 different input conditions of the SOP [21].

The performance of our all-optical polarization scrambler was first experimentally characterized by evaluating the degree of polarization (DOP) as a function of the backward power, i.e. the amplification coefficient $g$. The DOP is classically defined as $DOP = (\langle s_{L1}\rangle^2 + \langle s_{L2}\rangle^2 + \langle s_{L3}\rangle^2)^{1/2}$ and is used to quantify the scrambling of the output SOP $s(L,t)$ over the Poincaré sphere. As it can be seen in Fig. 5a (blue circles), the DOP of the output signal has initially a value close to unity, which is related to the constant-in-time output SOP that characterize the attraction regime.

The DOP starts to decrease beyond the amplification threshold $g\simeq 8$, in perfect agreement with our theoretical prediction of $g_A$. When $g>8$ the system enters in a transition regime. In such a regime small variations of the amplification $g$ could give rise to different temporal trajectories of the output SOP that cover only partially (see Fig. 4b) or almost entirely (see Fig. 4e) the Poincaré sphere. For this reason some fluctuations can be observed in the DOP function.

Finally, for high values of $g$, typically above $g_c = 5g_A = 40$, the system enters into the chaotic scrambling regime. The experimental DOP remains lower than 0.3, which corresponds to an efficient scrambling of the output SOP all over the Poincaré sphere. In particular, each sequence of $S_1$, $S_2$ and $S_3$ is characterized by autocorrelations that rapidly tend to zero, indicating that they don't exhibit deterministic repetitive patterns, in agreement with our numerical predictions. We stress here the fact that in this regime the Lyapunov coefficient $\lambda(z=L)$ is always positive (Fig. 5b), which provides a key signature of the chaotic nature of the dynamics of the output SOP.

Moreover, it is important to note that, contrary to the transition regime, in the chaotic regime the output SOP dynamics does not depend on the particular input SOP or particular rotation matrix $R$. For this reason in Fig. 5 when $g>g_C$ there are no noticeable differences between the 3 solid curves that refer to the numerical solution of Eqs. 1 with 3 different input SOPs and matrices $R$ which are chosen in a random fashion. The system also becomes independent of the input power, but in practice we have observed that the more the input power is, the easier the system enters into the chaotic regime. In fact, when the device operates with moderate powers, an adjustment of the polarization controller PC is needed in order to force the system to evolve into an unstable chaotic region.

In addition we have also checked that, for the typical values of power used in these experiments, no polarization scrambling occurs if the counter-propagating beam is an external wave generated independently of the forward wave [22, 23]. This clearly indicates that the instability of the system is in fact fundamentally related to the feedback effect imposed by the reflected loop setup.

The dynamics of the system is even more striking when

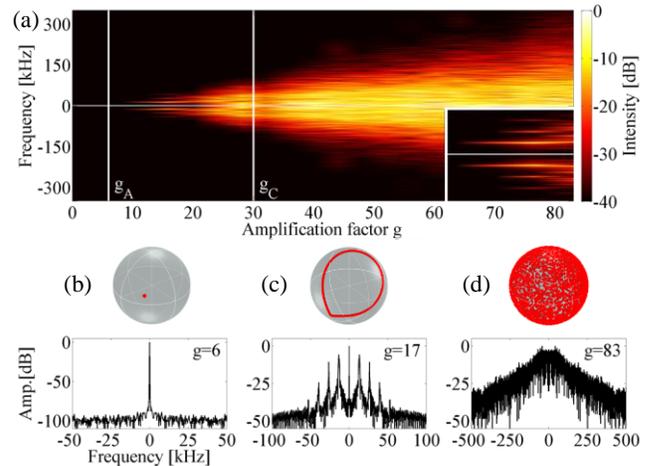

Fig. 6 (a) Experimental RF spectrum of the output component $S_1$ as a function of the amplification factor g. Snapshots (b), (c) and (d) illustrate the spectrum of $S_1$ and the output SOP distribution over the Poincaré sphere when the backward power is 20dBm ($g=4$), 23dBm ($g=8$) and 33dBm ($g=80$), respectively.

monitored in the spectral domain. For that purpose, we have measured the RF spectrum of the output Stokes component $S_1$. These measurements were achieved by recording the electrical spectrum behind a photodiode detecting the output beam through an optical polarizer.

Fig. 6a displays the evolution of the output RF spectrum as a function of the backward power $P_J(L)=gP_S(L)$ and the corresponding $g$ parameter. The 3 regimes previously discussed are distinctly visible:

- When $P_J(L)<23$ dBm ($g<g_A=8$), we are in the attraction regime: the spectrum always exhibits a single narrow peak centered in f=0 Hz, which corresponds to a constant-in-time value in the temporal domain.

- When 23 dBm<$P_J(L)$<30 dBm ($8<g<40$), the transition regime is reached, and as previously mentioned the system can exhibit 3 different dynamics: either an attraction towards an unstable stationary point, which corresponds to a DOP close to the unity (see in Fig. 5 the experimental results for $g\simeq20$) and to a narrow peak in the RF spectrum (see Fig. 6 for $g\simeq20$); either a periodic trajectory, which corresponds to equally spaced narrow peaks in the RF spectrum; or a semi-chaotic trajectory for which the spectrum begins to broad.

- Finally in the chaotic regime, for $P_J(L)>30$ dBm ($g>40$) the spectrum evolves in a much broader continuum of frequencies without showing any discrete component, which corresponds to an increasing scrambling speed and true chaotic behavior of the output polarization.

The snapshots (b-d) in Fig. 6 and corresponding Poincaré spheres illustrate the 3 typical regimes of our scrambler. Snapshot (b) for $P_J(L)=20$ dBm ($g=4$) depicts the attraction regime, characterized by a single peak centered in f=0 Hz in the RF spectrum of $S_1$. For $P_J(L)=23$ dBm ($g=8$), the snapshot (c) shows the transient regime: discrete frequency harmonic components in the RF spectrum are localized at n×13 kHz, to whom it corresponds a closed and periodic trajectory on the Poincaré sphere. Finally, snapshot (d) reports an example of the scrambling regime for $P_J(L)=33$ dBm ($g=80>g_c$); we can clearly see a continuum of frequencies in the RF spectrum, which is characterized by an almost uniform coverage of the surface of the sphere and thus an efficient polarization scrambling of the output signal. Note in this respect that a theoretical description of the spectral dynamics of the Stokes components in this chaotic regime of the polarization scrambler is in progress by making use of the wave turbulence theory [24].

Our home-made polarization scrambler was also tested for Telecom applications. In particular, we have characterized the degradations of the intensity profile due to the nonlinear regime undergone by the signal during the propagation. The initial incoherent signal was thus replaced by a 10-Gbit/s OOK signal centered at 1550-nm. The input SOP was kept constant and the injected power in Fibre-1 was fixed to $Ps(0)=15$ dBm. Figure 7a displays the output Poincaré sphere of the 10-Gbit/s signal for a backward power of 30 dBm ($g\simeq40$) and corresponds to an experimental scrambling speed of 107 krad/s. It thus confirms that an efficient scrambling process can be achieved, even with high-repetition rate pulsed signals. Moreover, Fig. 7b shows that the shape of the pulses is also remarkably preserved with a clear opened output eye-diagram, which validates the applicability of our polarization scrambler to RZ telecom signals.

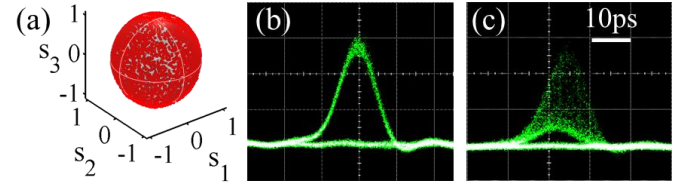

Fig. 7 Poincaré sphere (a) and eye-diagram (b) of the 10-Gbit/s signal recorded at the output of fiber-1 for an input average power of 15 dBm and a backward power of 30 dBm. Panel (c): output eye-diagram for a backward power of 35 dBm.

Finally, due to its intrinsic principle, the main limitation of our system lies in the strong Rayleigh back-scattering generated by the high-powered counter-propagating replica. In fact, the Rayleigh emission is coupled to the scrambled signal through the output circulator, which induces a non-negligible amount of noise at the signal frequency. This phenomenon is well illustrated in Fig. 7c, where the backward power is increased up to 35 dBm ($g\simeq120$). The corresponding output eye-diagram turns dramatically closed with a high level of amplitude jitter. As a consequence, this deleterious effect limits the maximum backward power that can be re-injected into the fiber and thus the scrambling speed that can be achieved. A practical solution to limit this drawback would be to use a frequency offset pump channel which first copropagates with the initial signal but still remains the only back-reflected beam. This technique has been implemented in the last section of the paper for the WDM configuration.

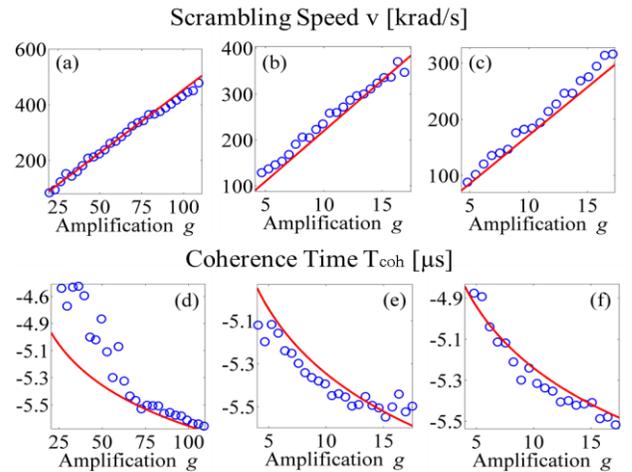

Fig. 8 Experimental measurements (blue circles) and analytical estimations by means of Eqs.(7-8) (red solid line) of the scrambling speed and coherence time. See text for details about the fiber parameters (Fibre-1,2) and the injected powers (P-1,2). Panels (a,d): case of Fibre-1 and P-1 (for which $g_C\simeq40$); panels (b,e): case of Fibre-1 and P-2 ($g_C\simeq8$); panels (c,f): case of Fibre-2 and P-2 ($g_C\simeq6$).

## VI. SCRAMBLING PERFORMANCES

In this section we analyze the performances of our device in terms of scrambling speed and coherence time. The scrambling speed $v$ represents the average angle covered by **s** in 1 second over the Poincaré sphere:

$$v = \langle |\partial_t \mathbf{s}| \rangle \quad (5)$$

The output coherence time is defined as $T_{coh}=(t_{c1}+t_{c2}+t_{c3})/3$, being $t_{ci}$ ($i=\{1,2,3\}$) the coherence time related to the component $s_{Li}$, that is the area of the associated auto-correlation function [21]:

$$t_{ci} = \int_{\tau=-\infty}^{\infty} \langle s_{Li}(t+\tau) \cdot s_{Li}(t) \rangle \quad (6)$$

This parameter reveals how fast the polarization fluctuations of the output SOP become uncorrelated, and is thus an important quantitative index to evaluate how quickly the depolarization process occurs.

In order to derive an analytical estimation of the output scrambling speed as a function of the system parameters, we have fitted numerical results by means of a least-square interpolation, using as model function $v_m = k_1 \cdot \gamma \cdot g \cdot P_S(0) \cdot 10^{-k_2 \cdot \alpha \cdot L}$. The best fitting to the numerical data is obtained when $k_1 = 4c_0/9$, where $c_0$ is the speed of light in the vacuum, and $k_2 = 1/6$. Furthermore, we found that the output coherence time is well interpolated by $1/v$, which leads to the estimations:

$$v = (4/9)c_0 \, \gamma \, gP_S(0)10^{-\alpha L/6} \quad (7)$$
$$T_{coh} = 1/v \quad (8)$$

The validity of these estimations is illustrated in Fig. 8, where the scrambling speed and the coherence time calculated by means of Eqs. (7-8) are in excellent agreement with the experimental measurements, obtained by means of the 100-GHz incoherent signal and both Fibre-1 and Fibre-2. Moreover, two input powers $P_s(0)$ were employed: P-1=15 dBm and P-2=22 dBm.

Analytical expressions of Eqs. (7-8) and results in Fig. 8 confirm the tendency previously observed in [12] namely that in the chaotic regime the scrambling speed grows up linearly with $g$. This confirms that $g$ is the key parameter that controls the temporal fluctuations of the output polarization. The scrambling speed reaches some hundreds of krad/s. Although this value is smaller than those of commercially available devices, it makes our chaotic scrambler of practical interest for the testing of real fiber optic systems. Note also that by increasing the fiber length, one can reduce the thresholds $g_A$ and $g_C$. The drawback is, however, that this also increases the total propagation losses, which degrades the scrambling performances and sets therefore a limit on the maximum admissible value of the fiber length $L$.

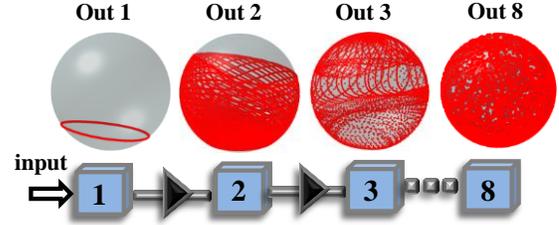

Fig. 9 Representation of the sequential cascade of 8 scramblers. See text for details on system parameters. The cyan numbered boxes identify the scramblers in the cascade; the black triangles indicate the re-amplification of the forward signal between two consecutive scramblers. The output SOP at the exit of the scramblers in the cascade is plotted over the Poincaré sphere when the amplification is set to $g=12$ in each scrambler.

## VII. CASCADE OF SCRAMBLERS

In order to overcome the limits imposed on the fiber length and on the backward power, discussed in the previous sections, we may take advantage of higher Kerr nonlinearities, e.g., in bismuth, tellurite, chalcogenide fibers, or more generally in soft-glass fibers [25-27], so as to make our polarization scrambler faster and compact. A different approach consists in implementing a cascade of scramblers where the forward beam **S** exiting the $n^{th}$-scrambler is amplified at its original power and then injected in the $(n+1)^{th}$-scrambler as illustrated in Fig. 9.

Actually, the cascade configuration permits to limit both the fiber length and the backward power at each stage, in such a way that propagation losses and back Rayleigh scattering remain negligible. For this reason high scrambling performances are guaranteed at each stage and, therefore, at the output of the cascade.

In order to achieve an efficient cascade effect, we should make sure that each stage operates in the chaotic regime (or at least in the transition regime), i.e., the amplification factor $g_n$ of the $n^{th}$-scrambler should be larger than the threshold $g_{A,n}$ related to the same scrambler.

In Fig. 9 the distribution over the Poincaré sphere of the SOP **s** at the output of the stages in a cascade of 8 scramblers is displayed. The fiber employed in each scrambler is Fibre-1 ($L$=5.3-km), thus the total length of the cascade system is $L_{cascade}$=42.4-km. The power $P_s(0)$ injected in the cascade is 15 dBm, and is kept constant at the entry of each stage thanks to the re-amplification of the forward signal between two consecutive scramblers. Therefore $g_{A,n} \equiv g_A \simeq 8$ for any scrambler of the cascade. The amplification gain $g_n$ is set to 12 at each stage, so that all scramblers work in the transition regime. In this configuration the SOP **s**, which follows a simple periodic trajectory at the output of the first scrambler (Out 1 in Fig. 9), becomes more and more scrambled and chaotic at the output of the following scramblers.

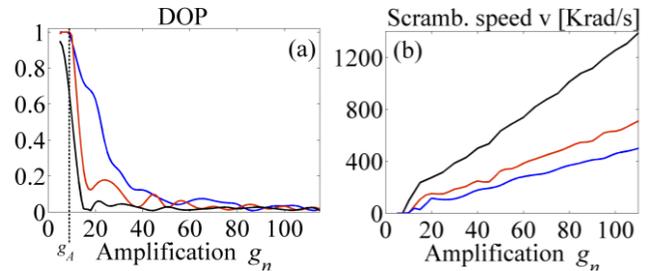

Fig. 10 Degree of polarization (DOP) and scrambling speed $v$ of the output SOP at the exit of the 1st scrambler (blue line), 2nd scrambler (red line) and 8th scrambler of the cascade that is represented in Fig.9. The amplification $g_n$ is the same at each stage. The black dotted vertical line indicates the cut-off $g_A$ for the DOP at the exit of cascade.

It is important to highlight that if an unique scrambler with a fiber of length $L_{cascade}$ were employed then the scrambling performances would be completely degraded by losses (see Eqs (7,8)). Despite losses, the implementation of the cascade process permits to achieve polarization fluctuations much faster. In Fig. 10, for the cascade of Fig. 9, we show the numerical calculation of the DOP and of the scrambling speed at the exit of consecutive stages as a function of the amplification $g_n$, which we assume to be the same in each scrambler.

We note that the scrambling speed is strongly improved at each stage, and that at the exit of the 8th stage the speed is increased up to four times with respect to the exit of the 1st scrambler. Furthermore, the DOP becomes lower at each stage, which indicates a uniform coverage of the whole Poincaré sphere at the exit of the cascade even with a low amplification gain $g_n$.

Our numerical simulations prove indeed that when using several scramblers in cascade, a sharp cut-off in proximity of $g_A$ is observed in the DOP function related to the exit of the last scrambler (see black line in Fig. 10a), so that if $g_n < g_A$ then the DOP is close to unity, while if $g_n > g_A$ then the DOP is close to zero. We thus infer that while in a single scrambler the threshold amplification $g_C$, is typically 5-10 times the threshold $g_A$, in a cascade configuration this threshold is reduced down to nearly $g_A$, that is to say $g_{C,cascade} \simeq g_A$. Moreover, the cascade allows for a considerable increment of the scrambling speed: we expect that by implementing a cascade of scramblers with highly-nonlinear fibers, e.g. $\gamma > 10$ $W^{-1}km^{-1}$), a speed of some Mrad/s could be reached.

## VIII. WDM CAPABILITIES

In this section, we experimentally characterize the behavior of our all-optical scrambler in the context of a wavelength division multiplexing (WDM) transmission. To this aim, we implement the experimental setup depicted in Fig.11. The initial signal first consists in 10-GHz short pulses generated from a mode-locked fiber laser (MLFL) at 1551 nm with 2-ps full width at half maximum (FWHM). The pulses are encoded at 10 Gbit/s in the OOK modulation format using a $2^{31}-1$ pseudo-random binary sequence (PRBS). The resulting data signal is then amplified to 30 dBm by means of an EDFA and injected into a 500-m long dispersion-flattened highly non-linear fiber (DF-HNLF from *ofs*) in order to broaden the spectrum through self-phase modulation and associated wave-breaking phenomenon [28]. The DF-HNLF is characterized by a chromatic dispersion of D=-1 ps/nm/km at 1550-nm, a dispersion slope of 0.006-ps²/nm/km, fiber losses of 0.6-dB/km and a nonlinear Kerr coefficient of 10.5-$W^{-1}.km^{-1}$. Five 10-Gbit/s OOK WDM channels and an additional pump channel are then sliced into the resulting continuum by means of a programmable optical filter (Waveshaper WS).

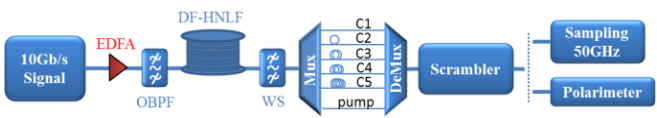

Fig. 11 Experimental setup for testing the chaotic polarization scrambler in WDM configuration. WS: Waveshaper.

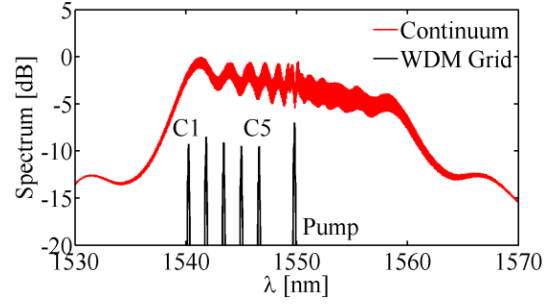

Fig. 12. Experimental continuum recorded at the output of the HNLF for an input power of 30 dBm (red). 10-Gbit/s WDM grid obtained by spectral slicing (black solid line).

As illustrated by Fig.12, which shows the experimental continuum recorded at the output of the HNLF and the resulting spectral grid, our final WDM signal consists in 5, 10-Gbit/s, 100-GHz spaced, 12-GHz bandwidth channels centered respectively at 1540,2 (C1), 1542 (C2), 1543,45 (C3), 1545(C4) and 1546,2-nm (C5), as well as a pump channel centered at 1550 nm.

All the WDM channels are then decorrelated in time into different polarization domains thanks to a combination of two optical demultiplexer/multiplexer with different delay-lines and polarization rotations for each channel before injection into our optical scrambler. As in the single-channel experiment described above, the 10-Gbit/s WDM signals is then injected into the system with a constant total average power of 15-dBm (7 dBm/channel). It is here important to notice that a 100-GHz optical bandpass filter was added into the reflective-loop so as to only keep the pump channel at 1550 nm for the backward signal whilst the 5 others WDM channels were picked out of the device for characterization. The role of this spectral routing operation was twofold: on the one hand, it ensures a unique state-of-polarization for the counter-propagating signal in order to maximize the efficiency of the scrambling process for all the transmitted channels, on the other hand it limits the deleterious impact of back Rayleigh scattering on the 5 other transmitted channels. At the output of the system, the 5 WDM channels were demultiplexed and individually characterized in polarization as well as in the time domain through the monitoring of the eye-diagram and bit-error-rate measurements.

Figures 13(a-c) display the Poincaré spheres of the different WDM channels recorded at the output of the all-optical scrambler. To not overload the paper, we report only 3 (among

5) Poincaré spheres representations, which correspond to C1: 1540.2 nm, C3: 1543.45 nm and C5: 1546.2 nm WDM channels, respectively; the 2 other channels exhibit similar performances. For this series of measurements, the total input power is kept constant to 15 dBm while only the 1550-nm pump channel is reflected and amplified in the backward direction with an average power of 29 dBm ($g$=31). Quite remarkably, we can first notice that despite the fact that the 5 input channels are initially uncorrelated so that each of them enters in the system with a different and unique SOP, the device is able to scramble the whole WDM grid. Indeed, in Fig.13 note that the SOP of each individual channel covers the whole surface of the Poincaré sphere and is characterized by a low value of its DOP, close to 0.2 for each channel.

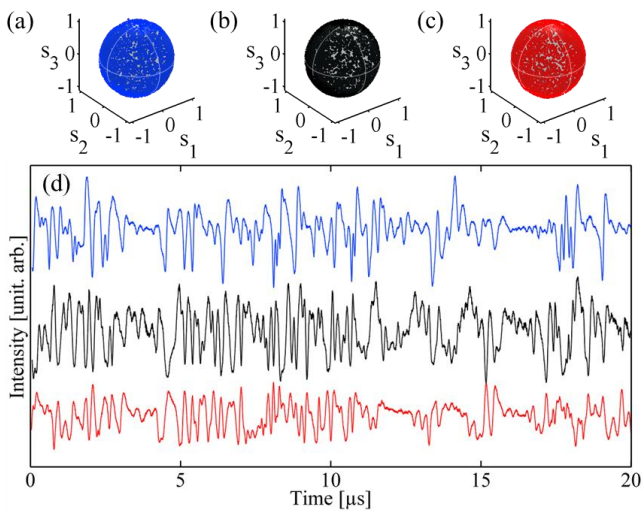

Fig. 13 (a-c) Output Poincaré spheres for WDM channels C1, C3 and C5, respectively. The input power is fixed to 15 dBm while the reflected 1550-nm pump channel is amplified to 29 dBm. (d) Intensity of the 3 output channels C1, C3 and C5 recorded behind a polarizer by means of a low bandwidth photodetector and oscilloscope.

It is interesting to note that, at the output of the device, all the random SOP trajectories undergone by the 5 different WDM channels are in fact correlated in time and are characterized by the same scrambling speed, close to 140 krad/s, see Table 1, in good agreement with results of numerical simulations. It is also important to notice that the scrambling speeds raised by all the WDM channels are roughly the same than the one measured in the previous single channel experiment. Indeed, this all-optical scrambler is mainly sensitive to the average power of the counter-propagative beam.

| | Scrambling speed (krad/s) | | | | |
|---|---|---|---|---|---|
| Channels (nm) | C1 1540.2 | C2 1542 | C3 1543.45 | C4 1545 | C5 1546.2 |
| Experiments | 156 | 120 | 132 | 112 | 114 |
| Numerics | 143 | 142 | 139 | 143 | 144 |

The time-correlation of the channels SOPs is also highlighted in Fig. 13d in which the intensity profiles of the 3 demultiplexed output WDM channels C1, C3 and C5 are synchronously recorded behind a polarizer by means of a low bandwidth photodetector and an oscilloscope. One can clearly notice the temporal correlation of polarization fluctuations between the different channels. This unexpected behavior can be intuitively interpreted considering the fact that in this configuration only one pump channel is back reflected, so that such pump imposes the polarization random walk on the Poincaré sphere of the others channels, as well as their scrambling speed.

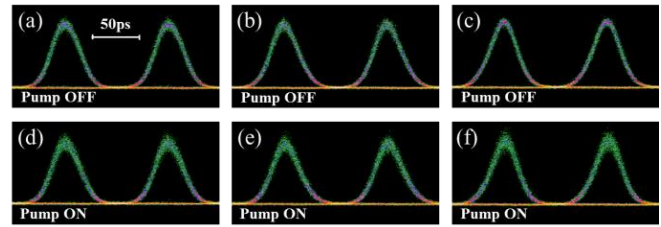

Fig. 14 (a-c) Output eye-diagrams in passive configuration (pump off) for WDM channels C1, C3 and C5, respectively; the input power is fixed to 15 dBm. (d-f) Corresponding eye-diagrams when the backward 1550-nm pump channel is amplified to 29 dBm.

The impact of the nonlinear polarization scrambling process on the 10-Gbit/s temporal profiles is illustrated in Fig. 14 for the WDM channels C1, C3 and C5. The other two channels have similar behavior. The upper row of insets underlines the high quality of the transmitted eye-diagrams obtained at the output of the fiber in passive configuration, i.e. when the backward 1550-nm pump channel is switched off. Note that in the WDM configuration the scrambling speeds are comparable with those of the single-channel configuration, but without increasing the total injected power at the system. This fact allows limiting the deleterious impacts of both self-phase and cross-phase modulation.

The bottom row shows the corresponding eye-diagrams when the backward pump channel is now switched on at an average power of 29 dBm, so that the scrambling process now operates efficiently.

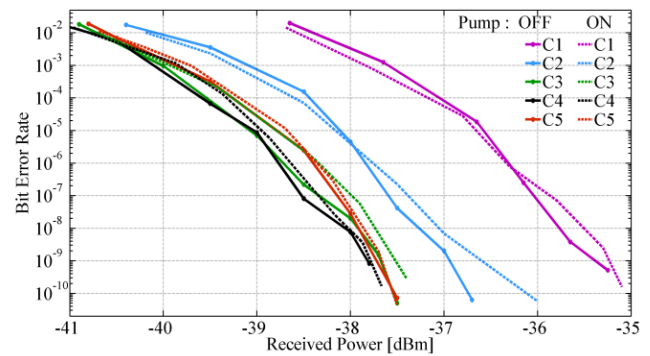

Fig. 15 Bit-error-rate measurements for the 5 WDM 10-Gbit/s channels recorded at the output of the nonlinear scrambler in passive configuration (pump off, solid lines) and in scrambling regime (pump on, circles). The input power is fixed to 15 dBm and the backward pump channel is set to 29 dBm.

Despite the high quality, wide opened eye-diagrams preserved by the nonlinear scrambling process, one can however observe a slight degradation of the temporal profiles with an increase of the amount of amplitude jitter. We attribute these impairments to the Rayleigh back-scattering provided by the spectrally broadened backward pump channel as well as a to weak Raman depletion effect caused by the pump to the signal. The impairments induced by the scrambling process on the 5 WDM 10-Gbit/s channels have been then quantified by means of systematic bit-error-rate measurements as a function of the received power in passive and pump on configurations, respectively. Results for the 5 WDM channels are summarized in Fig. 15 and show that a very weak power penalty is provided by the nonlinear scrambling process when comparing Pump ON/Pump OFF configurations. More precisely, a power penalty of 0.2 dB for the whole channels has been measured in average at a BER of $10^{-9}$.

CONCLUSIONS

In this work we have reported a theoretical, numerical and experimental description of an all-optical, fully chaotic nonlinear polarization scrambler.

The basic principle of this device was initially proposed in ref. [12]. It is based on the nonlinear cross-polarization interaction in a standard optical fiber between a forward signal and its high-power counter-propagating replica, generated and amplified by a factor $g$ at fiber end by means of a reflective loop setup. This system is in fact an extension of the device called Omnipolarizer [13] to a new chaotic operating regime. We gain here a deeper understanding of the physics underlying this all-optical scrambler. Indeed, we derive some useful analytical expressions of both the thresholds $g_A$ and $g_C$ that rule the transition between the different operating regimes of the device, as well as an analytical estimation of the scrambling speed and of the coherence time of the output polarization. These estimations are fully confirmed by numerical and experimental results, which draw the attention to the main factors that limit the scrambling performances.

In particular, experimental results obtained on a 10-Gbit/s OOK signal show that in a Telecom context our device is mainly limited by propagation losses and the detrimental Rayleigh back-scattering when large amplification gains $g$ are employed. These deleterious effects thus limit the scrambling speed of our system around 500 krad/s.

However, to overcome these drawbacks, another scenario has been also proposed and numerically studied which is based on a cascade of chaotic scramblers. This cascade of fibered scramblers allows to obtain an effective scrambling of the polarization even in presence of a relatively small amplification gain $g$, and could noticeably increase the output scrambling speed up to some Mrad/s, which is comparable to the speeds of the best commercially available systems.

Finally this nonlinear polarization scrambler has been also successfully tested in a 10-Gbit/s OOK WDM configuration. In particular, we have experimentally shown that this device is able to simultaneously scramble the polarization of 5 WDM channels and more surprisingly, despite its chaotic nature, it can impose a time-correlated random walk on the Poincaré sphere for each individual channel at an average speed close to 130 krad/s.

To conclude this home-made device, essentially based on standard components usually available in many Labs working in the field of nonlinear optics and optical communications, opens the path to the concept of fast and truly chaotic all-optical scrambling devices.

ACKNOWLEDGEMENTS

This research was funded by the European Research Council under Grant Agreement 306633, ERC PETAL. https://www.facebook.com/petal.inside. We also thank the financial support of the Conseil Régional de Bourgogne through the Photcom project. We thank Doc. S. Pitois for fruitful discussions.